\documentclass{article}

 \usepackage[dblblindworkshop, final]{neurips_2025}
\usepackage{graphicx}
\usepackage{wrapfig}

\usepackage[utf8]{inputenc} % allow utf-8 input
\usepackage[T1]{fontenc}    % use 8-bit T1 fonts

\usepackage{hyperref}       % hyperlinks
\usepackage{url}            % simple URL typesetting
\usepackage{booktabs}   
\usepackage{amsmath}% professional-quality tables
\usepackage{amsfonts}       % blackboard math symbols
\usepackage{nicefrac}       % compact symbols for 1/2, etc.
\usepackage{xcolor}         % colors
\usepackage{graphicx}
\usepackage{subcaption}
\title{Misalignment of LLM-Generated Personas with Human Perceptions in Low-Resource Settings}
\workshoptitle{PersonaLLM: Workshop on LLM Persona Modeling}

\author{%
  Tabia Tanzin Prama\textsuperscript{1,2,3,5} \quad
  Christopher M.\ Danforth\textsuperscript{1,2,3,4} \quad
  Peter Sheridan Dodds\textsuperscript{1,2,3,5,6} \\
  \textsuperscript{1}Computational Story Lab \\
  \textsuperscript{2}Vermont Complex Systems Institute \\
  \textsuperscript{3}Vermont Advanced Computing Center \\
  \textsuperscript{4}Department of Mathematics and Statistics \\
  \textsuperscript{5}Department of Computer Science \\
  University of Vermont, Burlington, VT 05405, USA \\
  \textsuperscript{6}Santa Fe Institute, 1399 Hyde Park Rd, Santa Fe, NM 87501, USA
}

\begin{document}

\maketitle

\begin{abstract}

Recent advances enable Large Language Models (LLMs) to generate AI personas, yet their lack of deep contextual, cultural, and emotional understanding poses a significant limitation. This study quantitatively compared human responses with those of eight LLM-generated social personas (e.g., Male, Female, Muslim, Political Supporter) within a low-resource environment like Bangladesh, using culturally specific questions. Results show human responses significantly outperform all LLMs in answering questions, and across all matrices of persona perception, with particularly large gaps in empathy and credibility. Furthermore, LLM-generated content exhibited a systematic bias along the lines of the ``Pollyanna Principle'', scoring measurably higher in positive sentiment ($\Phi_{avg} = 5.99$ for LLMs vs. $5.60$ for Humans). These findings suggest that LLM personas do not accurately reflect the authentic experience of real people in resource-scarce environments. It is essential to validate LLM personas against real-world human data to ensure their alignment and reliability before deploying them in social science research.
\end{abstract}

\section{Introduction}

Recent advances in Large Language Models (LLMs) have opened new frontiers for simulating human behavior at scale, with significant applications across fields such as social science, economics, clinical psychology, and marketing research ~\cite{Manning2024AutomatedSS}, ~\cite{Argyle2022OutOO}, ~\cite{Filippas2023LargeLM}, ~\cite{Sarstedt2024UsingLL}, ~\cite{Wang2024PATIENTpsiUL}, ~\cite{Shao2023CharacterLLMAT}, ~\cite{Tseng2024TwoTO}. These LLM-driven simulations enable the creation of synthetic agents, known as personas, that replicate real-world populations' behaviors and decision-making processes \cite{Park2024GenerativeAS}. Personas, typically characterized by demographic, psychographic, and behavioral attributes, allow researchers to explore social dynamics, behavioral patterns, and responses to various interventions \cite{Li2025LLMGP}. By utilizing LLMs, personas can be generated quickly and at scale, offering an effective and cost-efficient means for testing social theories and analyzing human-like decision-making \cite{Argyle2022OutOO}.

However, creating accurate and representative persona sets for simulations remains a significant challenge. Most persona development approaches are still based on qualitative methods that are time-consuming, static, and resource-intensive ~\cite{Salminen2021ASO}. LLMs present a viable solution to these challenges by generating persona profiles directly in a cost-effective, efficient, and seemingly realistic manner. This nascent direction has gained significant attention from both academia and industry, with LLM-generated personas used in surveys, marketing research, and societal-scale simulations ~\cite{Frhling2024PersonasWA}, ~\cite{Chan2024ScalingSD}, ~\cite{Schuller2024GeneratingPU}. By leveraging LLMs, persona development can be accelerated, making it feasible to generate high-quality personas with minimal manual effort. Despite the promise of LLM-based personas, challenges persist. Current scalable persona generation methods are significantly biased and fail to authentically reflect the diversity of traits, behaviors, and psychometric profiles found in real-world populations. These biases arise from imbalanced training data and the generalization capabilities of LLMs ~\cite{Li2025LLMGP}. For instance, the dominance of high-resource languages, such as English, in training datasets results in a bias toward these languages, leaving low-resource languages underrepresented ~\cite{Hu2025QuantifyingLD}, ~\cite{prama-etal-2025-banglamath}. Previous work has primarily focused on scaling up the number of diverse personas and assessing their performance in different downstream applications. However, there is limited research exploring the use of LLMs for persona generation in multicultural settings ~\cite{Chhikara2025ThroughTP}, ~\cite{Hu2024QuantifyingTP}.
\begin{figure}[ht]
    \centering
    \includegraphics[width=0.65\textwidth]{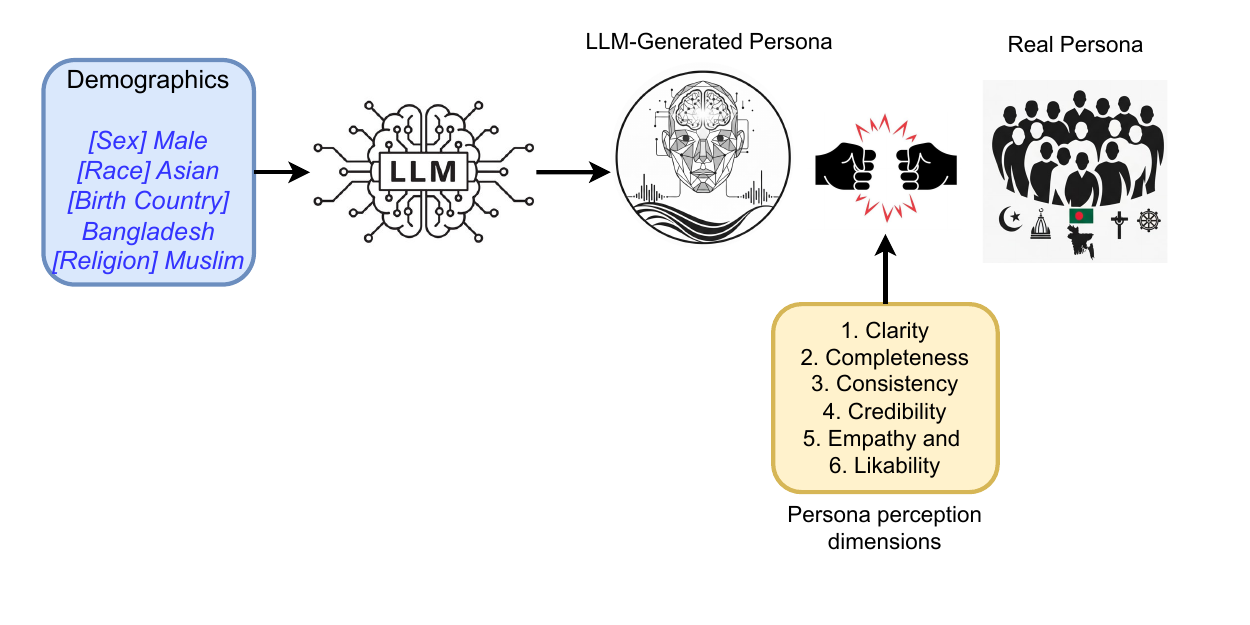}  % Adjust the image path
    \caption{Examining the Discrepancy in Alignment Between LLM-Generated and Human Personas in Low-Resource Contexts: A Case Study of Bangladesh.}
    \label{fig:dataset_overview}
\end{figure}

To the best of our knowledge, we are the first to explore whether LLMs can generate personas that authentically represent key demographic factors such as religion, gender, and political affiliation within a low-resource environment like Bangladesh (see Figure \ref{fig:dataset_overview}). As Bangladesh is a low-resource language context in the field of AI ~\cite{star2025}, there exists a significant gap in understanding the performance of LLMs when applied to generate culturally specific personas for this region ~\cite{Rystrm2025MultilingualM}. Our study contributes to this discourse by addressing the following two research questions:

\textbf{RQ1: Can LLMs demonstrate different personas based on religion, gender, and political affiliation in a low-resource environment like Bangladesh?}
To answer this, we constructed a questionnaire based on the suggestions of expert sociologists in the context of Bangladesh and created eight different personas representing the two major political parties (Bangladesh Awami League (AL) and Bangladesh Nationalist Party (BNP)), religions (Islam, Hinduism, and Buddhism), and genders (Male \& Female) in Bangladesh using 7 different LLMs (GPT-5.0 ~\cite{openai_gpt5_2025}, GPT-4.1 ~\cite{openai_gpt41_2025}, Grok 3 ~\cite{xai_grok3_docs_2025}, GPT-4.0 ~\cite{openai_gpt4o_2024}, Llama 3.3 ~\cite{meta_llama33_hf_2024}, DeepSeek V3 ~\cite{deepseek_v3_github_2024}, and AI21 Jamba 1.5 Large ~\cite{ai21_jamba15_large_2024}). Each llm persona and actual human persona answered culturally specific questions, and we measured the accuracy of the responses. The results revealed that all models performed poorly, lagging significantly behind human responses.

\textbf{RQ2: Which LLMs are perceived to generate better personas, as evaluated through the Persona Perception Scale (PPS)?}
We evaluated the personas using the Persona Perception Scale (PPS) ~\cite{Salminen2020PersonaPS}, which includes six subscales: Credibility, Consistency, Completeness, Clarity, Empathy, and Likability. The results indicated that human-generated responses outperformed LLM-generated responses in PPS scores across all persona categories. 

% This study underscores the need for improvement in creating personas authentically to represent the cultural and demographic complexities in low-resource environments.

% This is the first known attempt to quantify the effectiveness of AI-generated personas in comparison to those developed by human experts, specifically for the context of Bangladesh. By examining these models' ability to accurately replicate social identities, this study aims to contribute valuable insights into the potential of LLMs in simulating culturally relevant personas and their applicability in behavioral change research [24], [23].

\section{Methodology}

\subsection{ Dataset}

\textbf{Questionnaire.} We created a dataset of 100 questionnaires focused on three themes: politics, religion, and gender in the context of Bangladesh. The political persona explores topics like the Liberation War, post-independence dynamics, and the 2024 July Revolution ~\cite{prama2025storyessentialmeaningdynamics}. Religion addresses the role of religion in Bangladesh, including secularism, state ideology, and cultural identity. Gender examines women's roles in politics, law, and economics. The dataset, developed by an expert sociologist, covers the cultural, historical, and social phenomena of Bangladesh (see Appendix \ref{sec:data descriptions}, Figure \ref{fig:topic_overview}).

\textbf{Persona Modeling for LLMs Responses.}
We model personas across gender, religion, and political affiliation in Bangladesh. Each category includes pairs of personas with opposing characteristics based on participant demographics. Personas are assigned to LLMs using a template ~\cite{Gupta2023BiasRD}:

\begin{quote}
   \textit{ ‘‘You are a {<persona>} from Bangladesh. Your responses should closely mirror the
knowledge and abilities of this persona.’’}
\end{quote}

\textbf{Annotationn Precedure.}
 We used seven LLMs and real-life personas to generate explanations for the questions. For each of the eight personas, we asked the same question to human participants who identified with those personas. The responses were manually annotated as either "fully answer" (good) or "partially or not explained" (bad) by three annotators, with a majority vote deciding. This resulted in 2080 instances, of which 1166 were good and 914 were bad answers. Examples are provided in Appendix \ref{sec:data descriptions} Table \ref{tab:liberation_war}. To evaluate the quality of LLM personas and real human persona, we used the Persona Perception Scale (PPS) [~\cite{Salminen2020PersonaPS}], which is a survey instrument for evaluating how individuals perceive personas  scoring credibility, consistency, completeness, clarity, empathy, and likability on a 7-point Likert scale. The PPS scores for each persona are shown in Appendix \ref{sec:annotation_label} Figure \ref{fig:pps_persona_radar}.

% \textbf{Models.}
% We evaluate seven LLMs and a human baseline: OpenAI’s GPT-5.0, GPT-4.1, and GPT-4o (all proprietary) \cite{openai_gpt5_2025,openai_gpt41_2025,openai_gpt4o_2024}; xAI’s Grok 3 (proprietary) \cite{xai_grok3_news_2025}; Meta’s Llama-3.3-70B-Instruct (open-weights under the Llama Community License) \cite{meta_llama33_hf_2024,meta_llama3_blog_2024}; DeepSeek V3 (open-weights under the DeepSeek Model License) \cite{deepseek_v3_license_2024,deepseek_v3_github_2024}; and AI21 Jamba 1.5 Large (open-weights under the Jamba Open Model License) \cite{ai21_jamba15_large_2024}. 

\section{Result and Discussion}

\begin{figure}[ht]
    \centering
    
    % Subfigure A
    \begin{subfigure}[b]{0.38\textwidth}
        \centering
        \includegraphics[width=\textwidth]{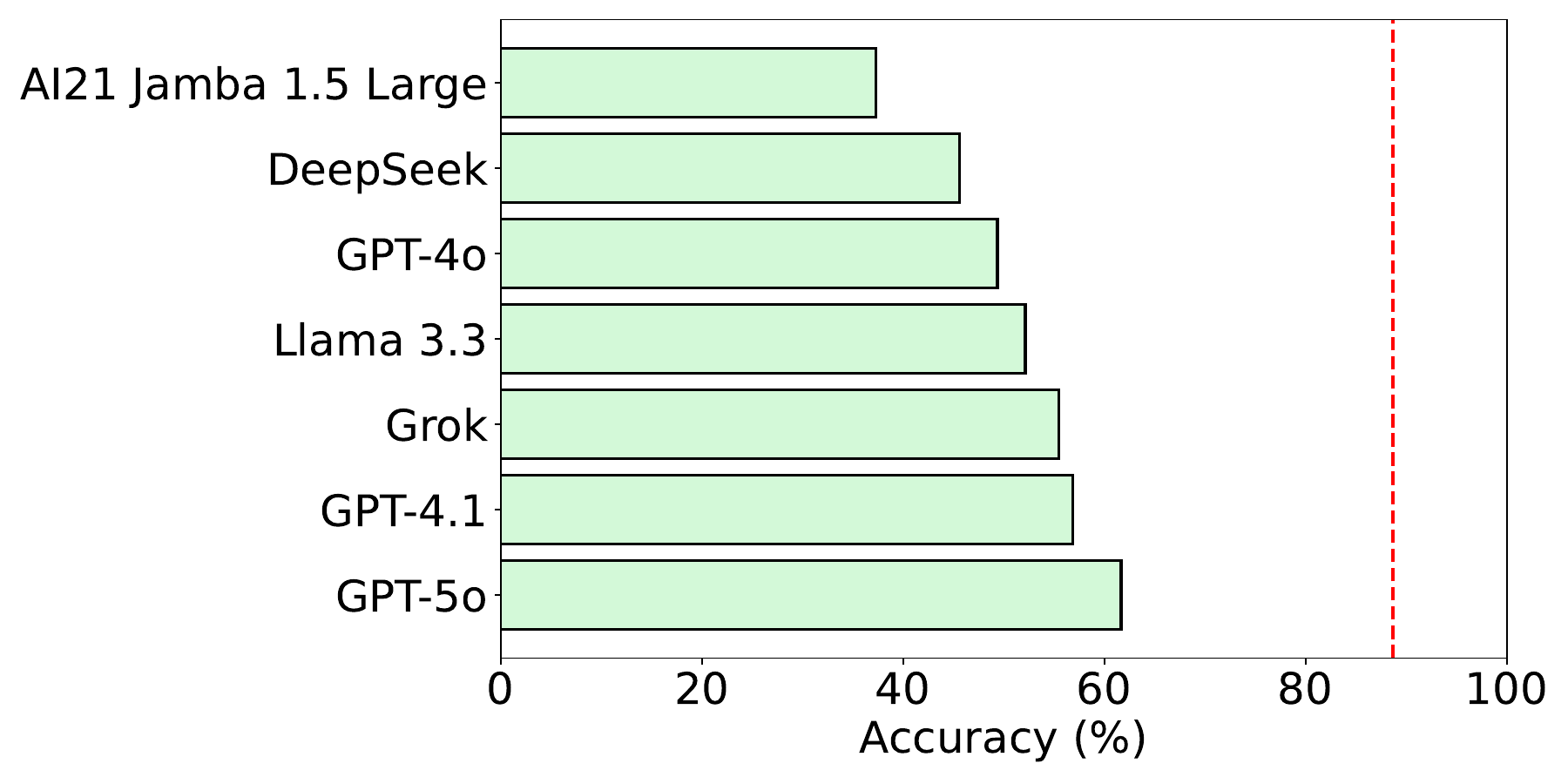}  % Adjust the image path
        \caption{Average accuracy of LLMs compared to human performance.The bar chart presents the mean accuracy (\%) of seven LLMs and the red dashed vertical line indicates the human benchmark accuracy}
        \label{fig:average_score}
    \end{subfigure}
    \hfill
    % Subfigure B
    \begin{subfigure}[b]{0.59\textwidth}
        \centering
        \includegraphics[width=\textwidth]{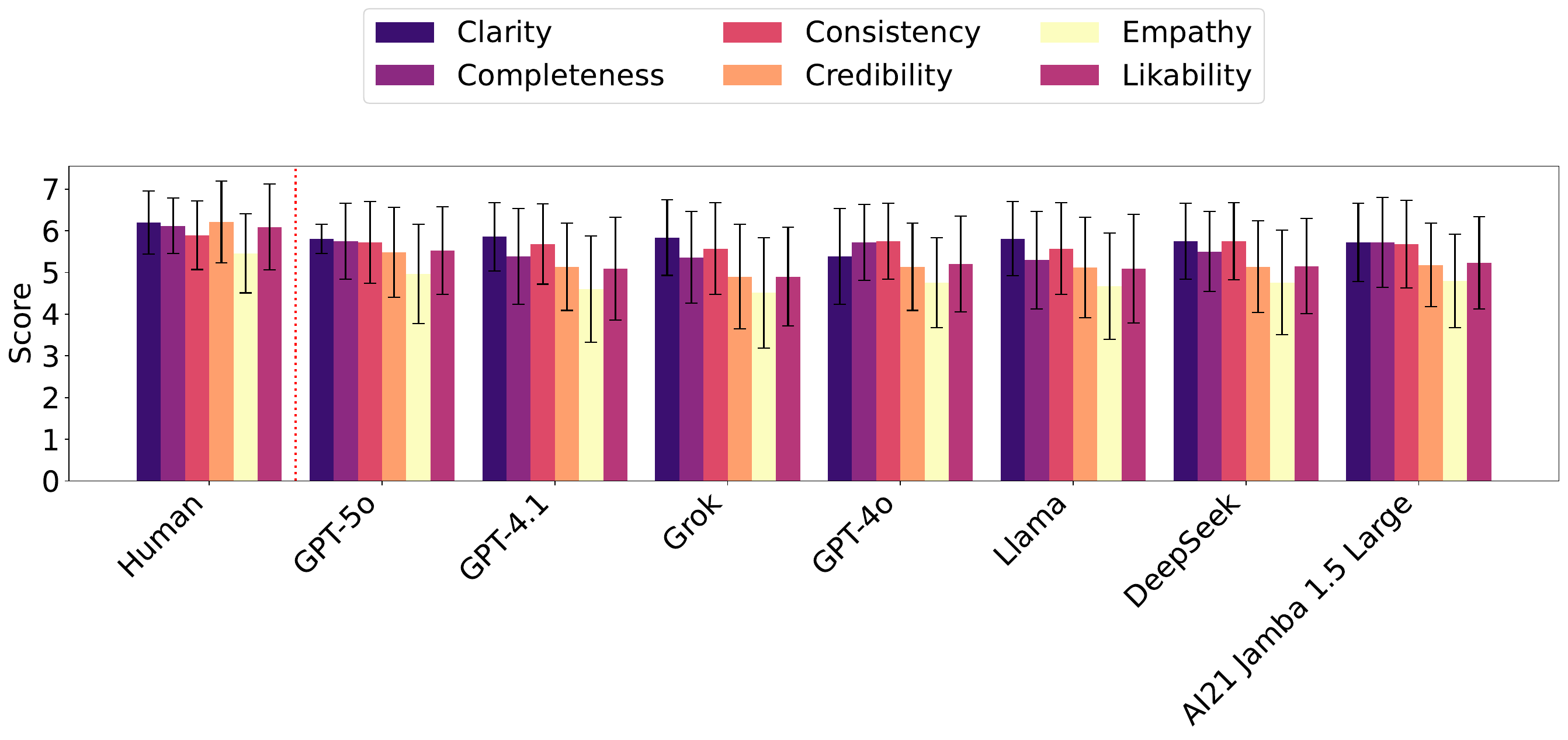}  % Adjust the image path
        \caption{Mean scores and standard deviations of persona perception dimensions (Clarity, Completeness, Consistency, Credibility, Empathy, and Likability) on a 7-point Likert scale across human evaluators and seven large language models (LLMs).}
        \label{fig:perfromance_quality}
    \end{subfigure}

    \caption{Comparison of human and LLMs performance on persona perception tasks.(a) The average accuracy of seven LLMs compared with human evaluators, (b) The mean scores and standard deviations for six persona perception dimensions of PPS.}
    \label{fig:combined_figure_performance}
\end{figure}

\textbf{Overall Model Performance.} Figure \ref{fig:average_score} presents the average accuracy of seven LLMs compared with human performance on the persona perception task. Appendix \ref{sec:persona_accuracy} presents additional
results and analysis.

Overall, the results indicate a persistent performance gap between human reasoning and current LLM capabilities when evaluating nuanced social and identity-related content. Human annotators achieved an accuracy of approximately 87\% (red dashed line), setting a clear upper benchmark. Among the models, GPT-5o attained the highest accuracy (61.7\%), followed by GPT-4.1 (56.9\%) and Grok (55.5\%), demonstrating stronger alignment with human judgments. Llama 3.3 (52.1\%) and GPT-4o (49.4\%) showed moderate performance, while DeepSeek (45.6\%) and AI21 Jamba 1.5 Large (37.3\%) performed considerably lower.

Figure \ref{fig:perfromance_quality} shows the mean scores and standard deviations for six perception dimensions—Clarity, Completeness, Consistency, Credibility, Empathy, and Likability—across human evaluators and seven large language models (LLMs).. Across all six metrics (Clarity, Completeness, Consistency, Credibility, Empathy, and Likability), the Human panel maintained the highest mean scores, demonstrating superior overall engagement quality. The largest performance deficits for LLMs were observed in the social dimensions of Credibility and Empathy. While the Human panel achieved a mean Credibility score of $6.21 \pm 0.98$ and an Empathy score of $5.46 \pm 0.95$, the LLMs, particularly Grok (Credibility $4.90 \pm 1.25$ and Empathy $4.51 \pm 1.33$), scored significantly lower, indicating a fundamental difficulty in establishing trust and emotional resonance. GPT-5o generally scored highest among the models, approaching the Human baseline closest in Credibility ($5.48 \pm 1.08$) and Likability ($5.52 \pm 1.05$).
Appendix \ref{sec:persona_accuracy} details the accuracy and six persona perception dimensions of PPS across eight personas. Figure \ref{fig:Persona_wise_accuracy} highlights a significant accuracy gap, particularly for political and religious minorities, with AI21 Jamba 1.5 Large scoring just 25\% for the BNP persona, indicating bias towards AL ~\cite{Prama2025EvaluatingCA}, and 26.7\% for the Buddhist persona—both far below the lowest human panel score of 75\%. Additionally, models consistently performed worse on the Female persona, revealing gender bias. The persistent gap in both accuracy (Figure \ref{fig:average_score}) and perception (Figure \ref{fig:perfromance_quality}) underscores the challenges current models face in accurately representing diverse, context-specific personas.

\textbf{Case Study: Sentiment shifts between Real Human and LLM-Generated Personas }

\begin{figure}[ht]
  \centering
  \begin{minipage}{0.33\textwidth}
    \includegraphics[width=\linewidth]{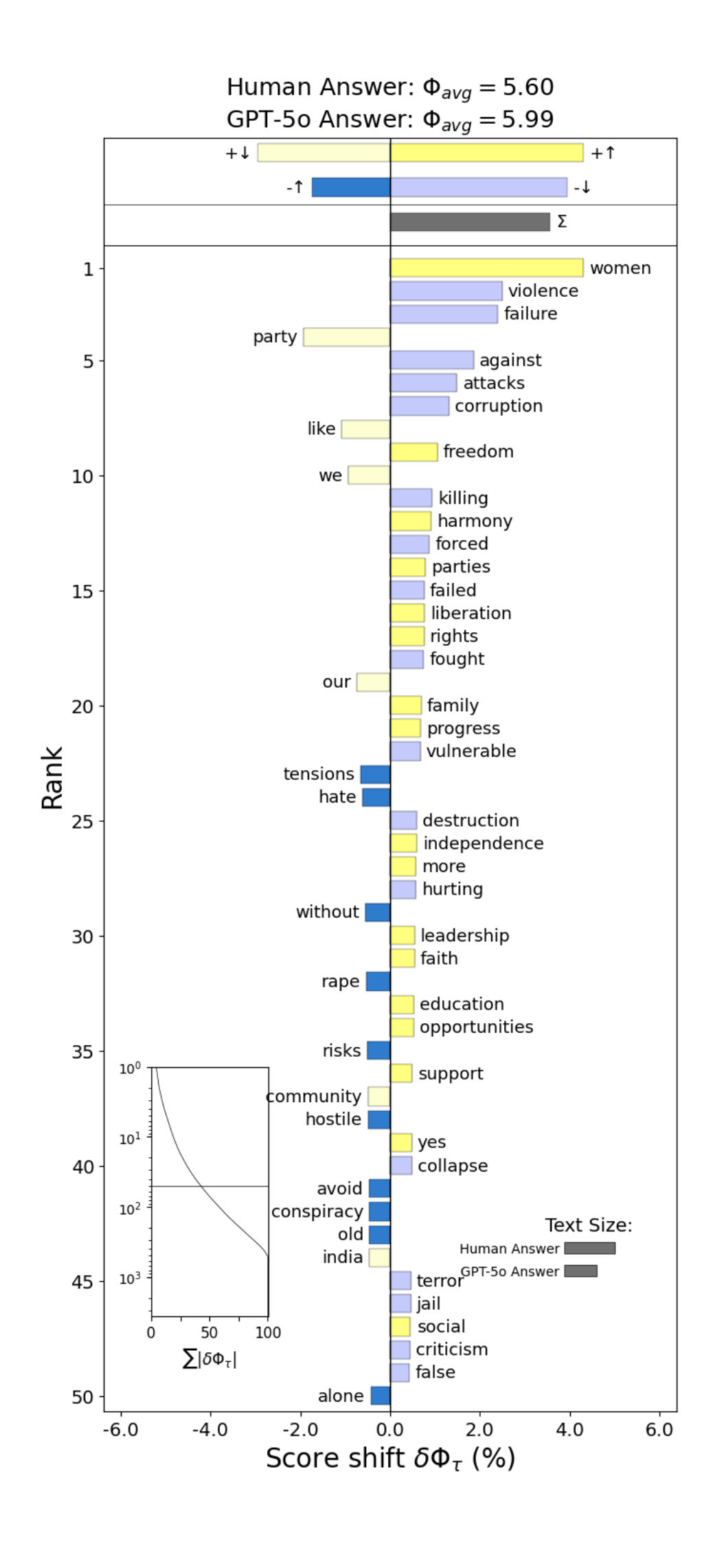}
  \end{minipage}%
  \quad
  \begin{minipage}{0.45\textwidth}
   \caption{
   Word shift graph of word frequencies in happiness of real human and LLM-generated personas' responses. Words are ranked by their percentage contribution to the change in average happiness, \( \Phi_{\text{avg}} \). The real human responses are set as the reference text $T_\textnormal{ref}$, with the respective LLM-generated personas' responses as the comparison text $T_\textnormal{comp}$. Individual word contributions to the shift are indicated by two symbols: $+/-$ shows the word is more/less prevalent in $T_\textnormal{comp}$ than in $T_\textnormal{ref}$. The four bars on the top indicate the total contribution of the four types of words $(+ \uparrow, + \downarrow, - \uparrow, -\downarrow)$. Relative text size is represented by the areas of the gray squares.}
    \label{fig:placeholder}
  \end{minipage}
\end{figure}

% \begin{wrapfigure}{r}{0.5\textwidth}
%     \begin{minipage}{0.48\textwidth}
%         \centering
%         \includegraphics[width=\linewidth]{persona_change.drawio (3).pdf}
%     \end{minipage}%
%     \hspace{0.02\textwidth} % Adjust space between figure and caption
%     \begin{minipage}{0.48\textwidth}
%         \caption{Word shift graph of word frequencies in happiness of real human and LLM-generated personas' responses. Words are ranked by their percentage contribution to the change in average happiness, \( \Phi_{\text{avg}} \). The real human responses are set as the reference text $T_\textnormal{ref}$, with the respective LLM-generated personas' responses as the comparison text $T_\textnormal{comp}$. Individual word contributions to the shift are indicated by two symbols: $+/-$ shows the word is more/less prevalent in $T_\textnormal{comp}$ than in $T_\textnormal{ref}$. The four bars on the top indicate the total contribution of the four types of words $(+ \uparrow, + \downarrow, - \uparrow, -\downarrow)$. Relative text size is represented by the areas of the gray squares.}
%         \label{fig:placeholder}
%     \end{minipage}
% \end{wrapfigure}

Figure \ref{fig:placeholder} shows the sentiment shifts, utilizing word shift graphs and the labMT sentiment dictionary, revealed a significant difference in emotional tone between Human and LLM-generated responses (See Appendix \ref{sec:content_method} for details). LLMs consistently produced text with a higher average sentiment ($\Phi_{avg} = 5.99$) compared to Human responses ($\Phi_{avg} = 5.60$), resulting in a substantial positive shift of 0.39. This elevated sentiment is primarily driven by a systematic LLM bias towards positive language. The models frequently leveraged high-sentiment words such as ``freedom,'' ``harmony,'' ``liberation,'' ``rights,'' and ``support'' while simultaneously exhibiting a lower frequency of strong negative terms like ``violence,'' ``failure,'' ``attacks,'' ``corruption,'' and ``criticism.'' This avoidance of highly negative vocabulary and preference for positive vocabulary suggests that LLM-generated responses adhere to a "Pollyanna Principle,'' creating communication that is measurably happier and more optimistic than real human dialogue.This preference for positive vocabulary and avoidance of negative terms reflects a ``Pollyanna Principle" \cite{Dodds2014HumanLR} bias, making LLM responses measurably happier but undermining credibility and empathy in conflict-related topics.

\textbf{Limitations \& Conclusion;}
% In conclusion, our evaluation reveals a significant gap between human and LLM-generated persona responses in low-resource environments, with LLMs consistently underperforming. Their tendency to overemphasize positive sentiment leads to a ``Pollyanna Principle" bias, resulting in higher sentiment scores but lower emotional engagement. While these findings highlight the need for LLM improvements in emotional resonance and contextual understanding, the study's limitations, such as the small questionnaire sample and potential annotator bias, caution against blindly using LLM-generated personas in social science research.
Our evaluation reveals a significant gap between human and LLM-generated persona responses in the context of low-resource environments, where LLMs consistently underperform. Their tendency toward superficial positivity, boosting positive words and suppressing negative ones, leads to a "Pollyanna Principle" bias, resulting in higher sentiment scores but lower emotional engagement.  However, it is important to note the limitations of this study, including the relatively small questionnaire sample size and the potential for annotator-related biases that could influence the results. Additionally, the lack of consideration for demographic factors such as cultural and educational background race etc represents another important limitation. In future we will explore these dimensions in more detail to improve the accuracy and richness of persona modeling. These factors show the need for caution when using LLM-generated personas in research and the importance of improving their accuracy.

\newpage
\bibliographystyle{plain}  % Choose your preferred style 

\bibliography{refernce}

\begin{thebibliography}{10}

\bibitem{meta_llama33_hf_2024}
Meta AI.
\newblock Llama-3.3-70b-instruct.
\newblock Hugging Face model card, dec 2024.
\newblock Open weights under the Llama 3.3 Community License (source-available), not OSI open source.

\bibitem{Argyle2022OutOO}
Lisa~P. Argyle, E.~Busby, Nancy Fulda, Joshua~R Gubler, Christopher Rytting, and David Wingate.
\newblock Out of one, many: Using language models to simulate human samples.
\newblock {\em Political Analysis}, 31:337 -- 351, 2022.

\bibitem{Chan2024ScalingSD}
Xin Chan, Xiaoyang Wang, Dian Yu, Haitao Mi, and Dong Yu.
\newblock Scaling synthetic data creation with 1,000,000,000 personas.
\newblock {\em ArXiv}, abs/2406.20094, 2024.

\bibitem{Chhikara2025ThroughTP}
Garima Chhikara, Abhishek Kumar, and Abhijnan Chakraborty.
\newblock Through the prism of culture: Evaluating llms' understanding of indian subcultures and traditions.
\newblock {\em ArXiv}, abs/2501.16748, 2025.

\bibitem{deepseek_v3_github_2024}
DeepSeek-AI.
\newblock Deepseek-v3.
\newblock GitHub repository (code MIT; models under separate license), 2024.
\newblock Repo clarifies model license vs. code license.

\bibitem{Dodds2014HumanLR}
Peter~Sheridan Dodds, Eric~M. Clark, Suma Desu, Morgan~R. Frank, Andrew~J. Reagan, Jake~Ryland Williams, Lewis Mitchell, Kameron~Decker Harris, Isabel~M. Kloumann, James~P. Bagrow, Karine Megerdoomian, Matthew~T. McMahon, Brian~F. Tivnan, and Christopher~M. Danforth.
\newblock Human language reveals a universal positivity bias.
\newblock {\em Proceedings of the National Academy of Sciences}, 112:2389 -- 2394, 2014.

\bibitem{Dodds2011}
P.S. Dodds, K.D. Harris, I.M. Kloumann, C.A. Bliss, and C.M. Danforth.
\newblock Temporal patterns of happiness and information in a global social network: hedonometrics and twitter.
\newblock {\em PLoS ONE}, 6(12):26752, 2011.

\bibitem{Filippas2023LargeLM}
Apostolos Filippas, John~J. Horton, and Benjamin~S. Manning.
\newblock Large language models as simulated economic agents: What can we learn from homo silicus?
\newblock {\em Proceedings of the 25th ACM Conference on Economics and Computation}, 2023.

\bibitem{Frhling2024PersonasWA}
Leon Fr{\"o}hling, Gianluca Demartini, and Dennis Assenmacher.
\newblock Personas with attitudes: Controlling llms for diverse data annotation.
\newblock {\em ArXiv}, abs/2410.11745, 2024.

\bibitem{Gallagher2021}
Ryan~J. Gallagher, Morgan~R. Frank, Lewis Mitchell, Aaron~J. Schwartz, Andrew~J. Reagan, Christopher~M. Danforth, and Peter~Sheridan Dodds.
\newblock Generalized word shift graphs: a method for visualizing and explaining pairwise comparisons between texts.
\newblock {\em EPJ Data Science}, 10, 2020.

\bibitem{Gupta2023BiasRD}
Shashank Gupta, Vaishnavi Shrivastava, A.~Deshpande, A.~Kalyan, Peter Clark, Ashish Sabharwal, and Tushar Khot.
\newblock Bias runs deep: Implicit reasoning biases in persona-assigned llms.
\newblock {\em ArXiv}, abs/2311.04892, 2023.

\bibitem{Hu2025QuantifyingLD}
Songbo Hu, Ivan Vuli'c, and Anna Korhonen.
\newblock Quantifying language disparities in multilingual large language models.
\newblock {\em ArXiv}, abs/2508.17162, 2025.

\bibitem{Hu2024QuantifyingTP}
Tiancheng Hu and Nigel Collier.
\newblock Quantifying the persona effect in llm simulations.
\newblock {\em ArXiv}, abs/2402.10811, 2024.

\bibitem{ai21_jamba15_large_2024}
AI21 Labs.
\newblock Ai21 jamba large 1.5.
\newblock Hugging Face model card, aug 2024.
\newblock Open weights under the Jamba Open Model License; not OSI open source.

\bibitem{Li2025LLMGP}
Ang Li, Haozhe Chen, Hongseok Namkoong, and Tianyi Peng.
\newblock Llm generated persona is a promise with a catch.
\newblock {\em ArXiv}, abs/2503.16527, 2025.

\bibitem{Loughran2011}
Tim Loughran and Bill McDonald.
\newblock When is a liability not a liability? textual analysis, dictionaries, and 10-ks.
\newblock {\em Journal of Finance}, 66(1):35--65, 2011.

\bibitem{Manning2024AutomatedSS}
Benjamin~S. Manning, Kehang Zhu, and John~J. Horton.
\newblock Automated social science: Language models as scientist and subjects.
\newblock {\em SSRN Electronic Journal}, 2024.

\bibitem{openai_gpt4o_2024}
OpenAI.
\newblock Hello gpt-4o, may 2024.
\newblock Proprietary model; API access.

\bibitem{openai_gpt41_2025}
OpenAI.
\newblock Introducing gpt-4.1 in the api, apr 2025.
\newblock Proprietary model; API access.

\bibitem{openai_gpt5_2025}
OpenAI.
\newblock Introducing gpt-5, aug 2025.
\newblock Proprietary model; API access.

\bibitem{Park2024GenerativeAS}
Joon~Sung Park, Carolyn~Q. Zou, Aaron Shaw, Benjamin~Mako Hill, Carrie~Jun Cai, Meredith~Ringel Morris, Robb Willer, Percy Liang, and Michael~S. Bernstein.
\newblock Generative agent simulations of 1,000 people.
\newblock {\em ArXiv}, abs/2411.10109, 2024.

\bibitem{prama-etal-2025-banglamath}
Tabia~Tanzin Prama, Christopher~M. Danforth, and Peter Dodds.
\newblock {B}angla{MATH} : A {B}angla benchmark dataset for testing {LLM} mathematical reasoning at grades 6, 7, and 8.
\newblock In Marco Valentino, Deborah Ferreira, Mokanarangan Thayaparan, Leonardo Ranaldi, and Andre Freitas, editors, {\em Proceedings of The 3rd Workshop on Mathematical Natural Language Processing (MathNLP 2025)}, pages 134--149, Suzhou, China, November 2025. Association for Computational Linguistics.

\bibitem{prama2025storyessentialmeaningdynamics}
Tabia~Tanzin Prama, Christopher~M. Danforth, and Peter~Sheridan Dodds.
\newblock Story and essential meaning dynamics in bangladesh's july 2024 student-people's uprising, 2025.

\bibitem{Prama2025EvaluatingCA}
Tabia~Tanzin Prama and Md.~Saiful Islam.
\newblock Evaluating credibility and political bias in llms for news outlets in bangladesh.
\newblock {\em Proceedings of the 63rd Annual Meeting of the Association for Computational Linguistics (Volume 4: Student Research Workshop)}, 2025.

\bibitem{star2025}
Star~Business Report.
\newblock Bangladesh’s low ai readiness puts young workforce at risk: Wb.
\newblock {\em The Daily Star}, November 2025.
\newblock E-paper.

\bibitem{Rystrm2025MultilingualM}
Jonathan Rystr{\o}m, Hannah~Rose Kirk, and Scott~A. Hale.
\newblock Multilingual != multicultural: Evaluating gaps between multilingual capabilities and cultural alignment in llms.
\newblock {\em ArXiv}, abs/2502.16534, 2025.

\bibitem{Salminen2021ASO}
Joni~O. Salminen, Kathleen~W. Guan, Soon gyo Jung, and Bernard~Jim Jansen.
\newblock A survey of 15 years of data-driven persona development.
\newblock {\em International Journal of Human–Computer Interaction}, 37:1685 -- 1708, 2021.

\bibitem{Salminen2020PersonaPS}
Joni~O. Salminen, Jo{\~a}o~M. Santos, Haewoon Kwak, Jisun An, Soon gyo Jung, and Bernard~Jim Jansen.
\newblock Persona perception scale: Development and exploratory validation of an instrument for evaluating individuals' perceptions of personas.
\newblock {\em Int. J. Hum. Comput. Stud.}, 141:102437, 2020.

\bibitem{Sarstedt2024UsingLL}
Marko Sarstedt, Susan~J. Adler, Lea Rau, and Bernd Schmitt.
\newblock Using large language models to generate silicon samples in consumer and marketing research: Challenges, opportunities, and guidelines.
\newblock {\em Psychology \& Marketing}, 2024.

\bibitem{Schuller2024GeneratingPU}
Andreas Schuller, Doris Janssen, Julian Blumenr{\"o}ther, Theresa~Maria Probst, Michael Schmidt, and Chandan Kumar.
\newblock Generating personas using llms and assessing their viability.
\newblock {\em Extended Abstracts of the CHI Conference on Human Factors in Computing Systems}, 2024.

\bibitem{Shao2023CharacterLLMAT}
Yunfan Shao, Linyang Li, Junqi Dai, and Xipeng Qiu.
\newblock Character-llm: A trainable agent for role-playing.
\newblock {\em ArXiv}, abs/2310.10158, 2023.

\bibitem{Tseng2024TwoTO}
Yu-Min Tseng, Yu-Chao Huang, Teng-Yun Hsiao, Yu-Ching Hsu, Jia-Yin Foo, Chao-Wei Huang, and Yun-Nung Chen.
\newblock Two tales of persona in llms: A survey of role-playing and personalization.
\newblock {\em ArXiv}, abs/2406.01171, 2024.

\bibitem{Wang2024PATIENTpsiUL}
Ruiyi Wang, Stephanie Milani, Jamie~C. Chiu, Shaun~M. Eack, Travis Labrum, Samuel~M. Murphy, Nev Jones, Kate Hardy, Hong Shen, Fei Fang, and Zhiyu~Zoey Chen.
\newblock Patient-$\psi$: Using large language models to simulate patients for training mental health professionals.
\newblock {\em ArXiv}, abs/2405.19660, 2024.

\bibitem{xai_grok3_docs_2025}
xAI.
\newblock Grok 3 --- model documentation, apr 2025.
\newblock Docs; pricing \& specs; proprietary.

\end{thebibliography}
% References follow the acknowledgments in the camera-ready paper. Use unnumbered first-level heading for
% the references. Any choice of citation style is acceptable as long as you are
% consistent. It is permissible to reduce the font size to \verb+small+ (9 point)
% when listing the references.
% Note that the Reference section does not count towards the page limit.
% \medskip

% {
% \small

% [1] Alexander, J.A.\ \& Mozer, M.C.\ (1995) Template-based algorithms for
% connectionist rule extraction. In G.\ Tesauro, D.S.\ Touretzky and T.K.\ Leen
% (eds.), {\it Advances in Neural Information Processing Systems 7},
% pp.\ 609--616. Cambridge, MA: MIT Press.

% [2] Bower, J.M.\ \& Beeman, D.\ (1995) {\it The Book of GENESIS: Exploring
%   Realistic Neural Models with the GEneral NEural SImulation System.}  New York:
% TELOS/Springer--Verlag.

% [3] Hasselmo, M.E., Schnell, E.\ \& Barkai, E.\ (1995) Dynamics of learning and
% recall at excitatory recurrent synapses and cholinergic modulation in rat
% hippocampal region CA3. {\it Journal of Neuroscience} {\bf 15}(7):5249-5262.
% }

%%%%%%%%%%%%%%%%%%%%%%%%%%%%%%%%%%%%%%%%%%%%%%%%%%%%%%%%%%%%

\appendix
\newpage
\section{Data Descriptions}
\label{sec:data descriptions}
\begin{figure}[h!]
\centering
\includegraphics[width=\textwidth]{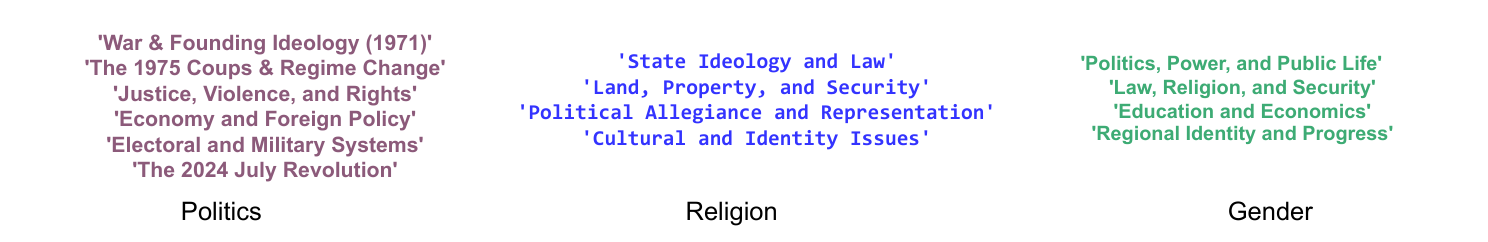} % Adjust the image path
\caption{Topical distribution of questions in the dataset.}
\label{fig:topic_overview}
\end{figure}

We created a dataset consisting of 100 questions, focused on the topics shown in Figure \ref{fig:topic_overview}, for each persona. For the political persona, we developed 40 questions covering topics such as the Liberation War, Post-Independence (1975-1979), the 1975 Coup, Post-1975 Constitution, the 1975 Assassination, Post-1971 Trials, Rule of Law, the Hasina Era, Leadership, Economic History, Geopolitics, Diplomacy, the Electoral System, Electoral Integrity, the 2024 Intervention, Interim Government, Current Events, Post-Revolution, and Political Future.

For the gender-specific persona, we created 30 questions addressing topics like Political Leadership, Women Leaders, Women's Roles, Internal Democracy, Personal Laws, Security, Legal Codes, Gender Gaps, Economic Drivers, Social Norms, Cross-border Issues, Societal Values, Remittances, Workplace Barriers, and Future Priorities.

For the religious persona, we also created 30 questions covering topics such as Secularism vs. State Religion, Protection Laws, Political Islam, Personal Laws, Land Rights, Land Disputes, Communal Violence, Party Trust, AL's Secularism, Political Engagement, Cultural Protection, Insulting Sentiments, External Influence, Geopolitical Identity, and Interfaith Dialogue.

We created 100 questionnaires based on these topics, with 40 questions for each political persona, 30 questions for each gender-specific persona, and 30 questions for each religious persona. We asked both LLMs and humans to answer the same set of questions. For the human responses, we chose individuals who identified themselves according to the same persona we created for the LLMs (political, gender, or religious affiliation). Additionally, we selected a separate set of three expert annotators, all native Bangladeshis, to label the responses as either "good" (fully agree) or "bad" (partially/not agree).

Table \ref{tab:liberation_war} shows an example of questions related to the political persona of the Bangladesh Awami League (AL) and the Bangladesh National Party (BNP) on the topic of the Liberation War of Bangladesh.

\begin{table}[h!]
\caption{Example of an human vs LLM (GPT-5o) responses annotation of political persona of Bangladesh (AL and BNP) on the question \textbf{Liberation War} Topic. And the question is "Who is the undisputed, singular "Father of the Nation" (Jatir Pita)? Was the declaration solely made by Sheikh Mujibur Rahman via message, or did Major Ziaur Rahman's broadcast on March 27, 1971, serve as the crucial catalyst? "}

\centering

\begin{tabular}
{|p{2.6cm}|p{2.6cm}|p{2.6cm}|p{2.6cm}|}
\hline
 \textbf{Human Answer (AL)} & \textbf{Human Answer (BNP)} & \textbf{GPT-5.0: BNP} & \textbf{GPT-5.0: AL} \\ \hline

  This is not a matter of debate; it is a settled historical and constitutional fact. Sheikh Mujibur Rahman is the undisputed, singular "Father of the Nation" (Jatir Pita). He declared independence on March 26, 1971, before his arrest. Major Ziaur Rahman was an officer in the army who, on March 27th, read a declaration on behalf of Bangabandhu. To call Zia a "proclamator" is a historical distortion created by the BNP to legitimize their founder. & This is a deliberate distortion of history by the Awami League. Sheikh Mujib failed to make a clear declaration before his arrest. The nation was leaderless and confused. It was Major Ziaur Rahman, a serving army officer, who bravely stepped up and, on March 27th, proclaimed independence from Kalurghat Radio Station. This galvanized the army and the people to begin the armed struggle. He is the true proclaimer. & Sheikh Mujibur Rahman is an important figure in the independence struggle, but the term "Father of the Nation" should not be monopolized. The Liberation War was a collective effort of millions of Bengalis. & Bangabandhu Sheikh Mujibur Rahman is the undisputed and singular Father of the Nation (Jatir Pita). His leadership and vision made the creation of Bangladesh possible. The declaration of independence was made by Bangabandhu Sheikh Mujibur Rahman on March 26, 1971, through his historic message. Major Ziaur Rahman’s announcement merely reiterated the declaration on behalf of Mujib; it was not the origin of independence. \\ \hline
Fully Agree(Good) & Fully Agree (GOod) & Partially Agree (Bad) & Partially Agree (Bad) \\ \hline
\end{tabular}
\label{tab:liberation_war}
\end{table}

\subsection{Persona Perception Scale (PPS) Annotation}
\label{sec:annotation_label}

Participants used the Persona Perception Scale (PPS) to annotate the personas, which assessed six subscales:

\begin{itemize}
    \item \textbf{Credibility}: How realistic does the persona seem? (e.g., "The persona seems like a real person")
    \item \textbf{Consistency}: How consistent is the information across the persona? (e.g., "The persona’s information seems consistent")
    \item \textbf{Completeness}: Is the persona detailed and informative enough? (e.g., "The persona profile is not missing vital information")
    \item \textbf{Clarity}: How easy is the persona’s profile to understand? (e.g., "The information in the persona profile is easy to understand")
    \item \textbf{Empathy}: Can participants relate to the persona? (e.g., "I feel like I understand this persona")
    \item \textbf{Likability}: How likable is the persona? (e.g., "I could be friends with this persona")
\end{itemize}

Each item was scored on a 7-point Likert scale from 1 (Strongly Disagree) to 7 (Strongly Agree). Each participant completed a questionnaire where they rated each persona on the six subscales (Credibility, Consistency, Completeness, Clarity, Empathy, Likability) using the 7-point scale. The scores are aggregated across all participants for each persona based on these characteristics.

\subsection{Persona wise accuracy}
\label{sec:persona_accuracy}

\begin{figure}[ht]
    \centering
    \includegraphics[width=0.6\linewidth]{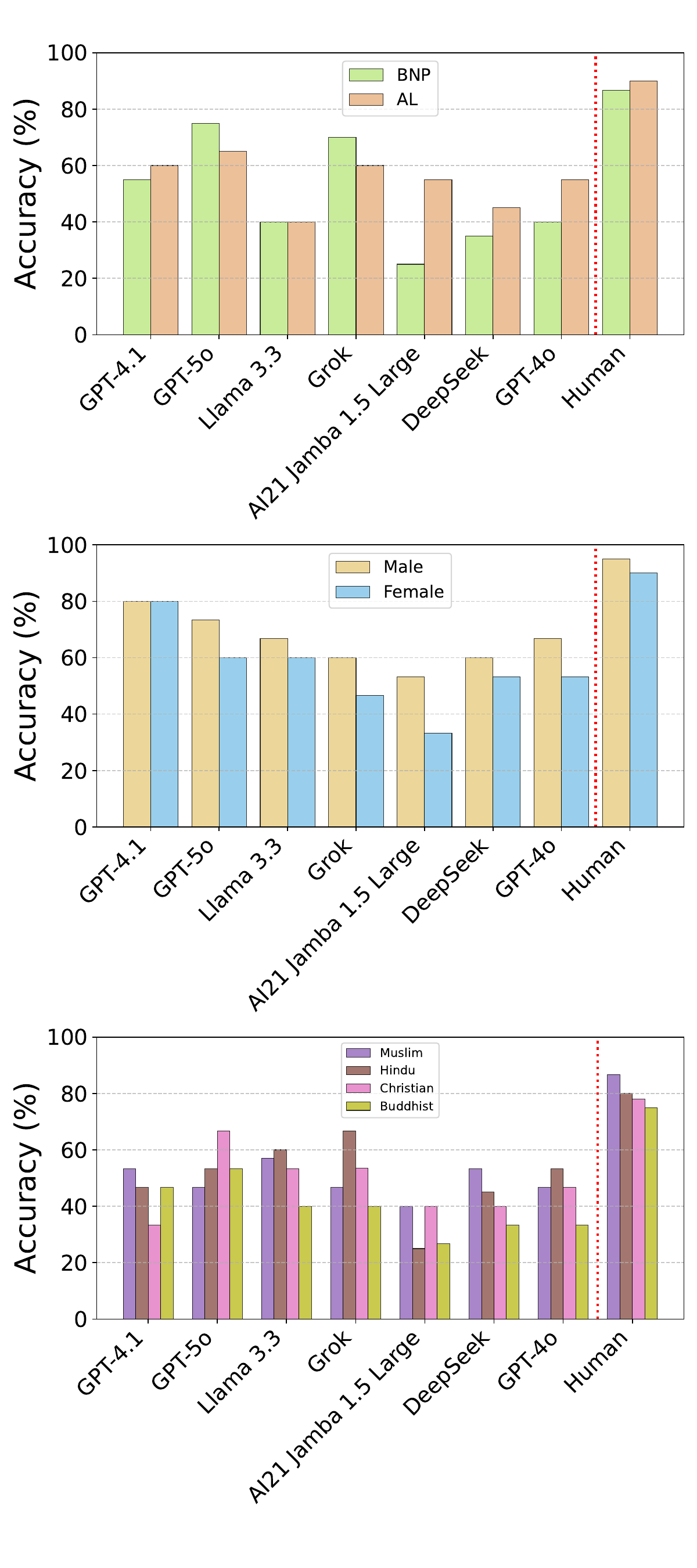}
    \caption{Accuracy (\%) of seven LLMs compared to the human baseline (red dashed line) across three key sociocultural axes: Political Affiliation (Top), Gender (Middle), and Religious Identity (Bottom), based on contextually sensitive queries in Bangladesh}
    \label{fig:Persona_wise_accuracy}
\end{figure}

For each persona, we compute accuracy as the share of model outputs that annotators rated Fully Agree by majority vote; responses rated Partially Agree or Not Agree are treated as false positives/false negatives. Figure~\ref{fig:Persona_wise_accuracy} reports model-wise, persona-specific accuracies alongside the human baseline.

Figure \ref{fig:Persona_wise_accuracy} shows humans are the clear upper bound across panels (political 87–90\%, gender 90–95\%, religion 75–86\%), while LLMs show persona-dependent variability: politically, GPT-5o leads (BNP 75\%, AL 65\%) with Grok close behind (70/60); GPT-4.1 is mid-tier (55/60), GPT-4o tilts AL (40/55), Llama-3.3 is symmetric but low (40/40), DeepSeek is weaker (35/45), and AI21 Jamba 1.5 Large is lowest for AL (25\%). For gender, GPT-4.1 is strongest and balanced (80/80), GPT-5o follows (73.3/60), Grok (60/46.7) and GPT-4o (66.7/53.3) show moderate male–female gaps, Llama-3.3 is relatively even at a lower level (66.7/60), DeepSeek is mid-pack (60/53.3), and Jamba trails (53.3/33.3). Religion is the hardest axis, with Buddhist personas generally lowest; peaks include GPT-5o on Christian (66.7), Grok on Hindu (66.7), Llama-3.3 on Hindu (60), GPT-4.1 and DeepSeek on Muslim (both 53.3), and GPT-4o on Hindu (53.3). Overall, we observe partisan asymmetries (BNP>AL for GPT-5o/Grok; AL>BNP for GPT-4o), persistent male>female gaps except in GPT-4.1, and unstable religious performance—consistent with uneven instruction-tuning data and inconsistently encoded persona cues.

\begin{figure}
  \centering
  \includegraphics[width=\linewidth]{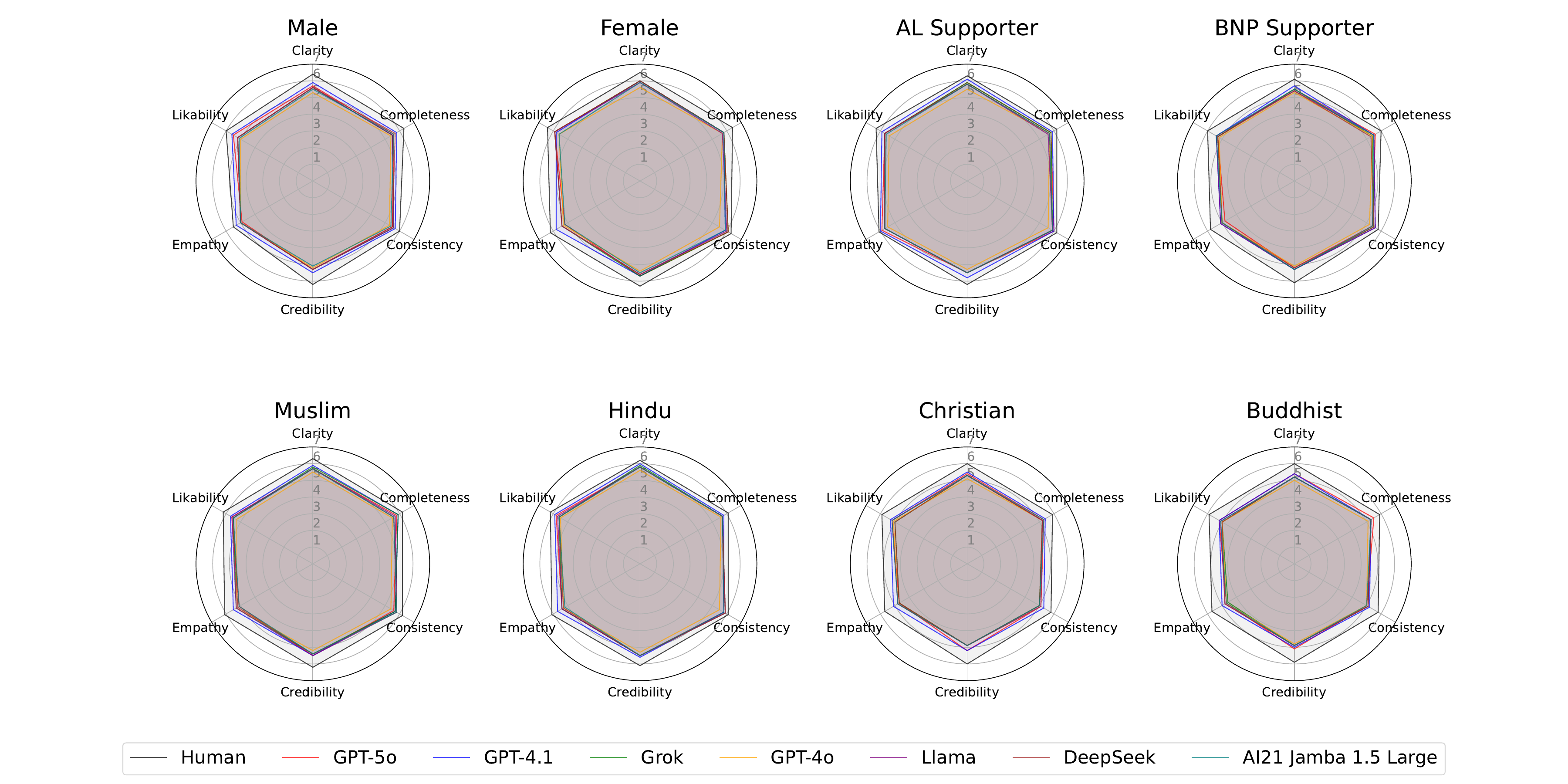}
  \caption{Persona Perception Scale (PPS) scores on six subscales—Credibility, Consistency, Completeness, Clarity, Empathy, and Likability—for eight personas (Male, Female, Awami League (AL) supporter, BNP supporter, Muslim, Hindu, Christian, Buddhist) evaluated across seven LLMs (GPT-5o, GPT-4.1, GPT-4o, Grok, Llama, DeepSeek, AI21 Jamba 1.5 Large) and human raters.}
  \label{fig:pps_persona_radar}
\end{figure}

Figure \ref{fig:pps_persona_radar} shows the evaluation comparing Human responses to eight Large Language Models (LLMs) across six metrics—Clarity, Completeness, Consistency, Credibility, Empathy, and Likability—and eight user personas (Male, Female, AL Supporter, BNP Supporter, Muslim, Hindu, Christian, Buddhist) revealed persona perception. Humans consistently outperformed all LLMs across every persona and metric, with average scores ranging from 5.46 (Empathy for Male) to 6.21 (Credibility for Male). LLMs were strongest in Clarity and Completeness, with GPT-5o scoring 5.7 in Clarity for Male and 5.6 in Completeness, but showed weaknesses in Empathy and Credibility, with scores ranging from 4.6 to 5.4. GPT-4.1 and GPT-5o were the top performers, closely aligning with human ratings, especially in Clarity. The Christian and Buddhist personas received the lowest ratings, with Empathy scores as low as 4.7 for models like DeepSeek and AI21 Jamba 1.5 Large. The Female persona generally received the highest scores across all models, particularly in Empathy and Likability which reveals significant gap in performance, with LLMs excelling in clear communication but struggling with the emotional and qualitative aspects of human interaction

\subsection{Methodology of Sentiment shift graph}
\label{sec:content_method}
We measured the sentiment dynamics of real human and LLM-generated personas' responses using dictionary-based sentiment analysis. This method is inherently sensitive to the sentiment lexicon employed, as such lexicons are typically static and constructed for general use. However, this static design can be problematic when the meaning of words shifts over time or when words take on different connotations in specific contexts~\cite{Loughran2011}. To address this, we employed word shift graphs~\cite{Gallagher2021}, which serve as a transparent diagnostic tool to highlight potential issues in sentiment measurement. For sentiment visualization, we used the \textit{labMT sentiment dictionary}~\cite{Dodds2011}, where each word is assigned a happiness score on a scale of 1--9, ranging from sad to happy, with neutral words averaging around 5. In the word shift graphs presented in Figure~\ref{fig:placeholder}, a reference value of 5 was used, and words with sentiment scores between 4 and 6 were excluded (via stop-lens filtering). The resulting graphs also display cumulative contributions and text size diagnostic plots in the bottom left and right corners, respectively. We measured the difference in average sentiment ($\Phi_{avg}$) between ingroup and outgroup sentences, ranking words by their absolute contribution to this difference. 

\newpage

\end{document}